\documentclass[aps,prl,reprint,superscriptaddress]{revtex4-1}

\usepackage{bm}% bold math
\usepackage{color}% text color
\usepackage{graphicx}
\usepackage{lastpage}
\usepackage{float}
\usepackage{fancyhdr}
\usepackage{fnpos}
\usepackage[english]{babel}
\usepackage{array}
\usepackage{droidsans}
\usepackage{charter}
\usepackage[T1]{fontenc}
\usepackage[usenames,dvipsnames]{xcolor}
\usepackage{setspace}
\usepackage[compact]{titlesec}
\usepackage{textcomp}
\usepackage{soul}

\makeatletter 
\newlength{\figrulesep} 
\makeatother
\begin{document}

\title{Probing the in-mouth texture perception with a biomimetic tongue}

\author{Jean-Baptiste Thomazo}
 \affiliation{Sorbonne Universit\'e, Centre National de la Recherche Scientifique, Laboratoire Jean~Perrin, LJP, F-75005 Paris, France.}

\author{Javier Contreras Pastenes}
 \affiliation{Sorbonne Universit\'e, Centre National de la Recherche Scientifique, Laboratoire Jean~Perrin, LJP, F-75005 Paris, France.}

\author{Christopher Pipe}
 \affiliation{Nestl\'{e} Research Center, Route du Jaurat, 1000, Lausanne, Switzerland}

\author{Benjamin Le R\'{e}v\'{e}rend}
 \affiliation{Nestl\'{e} Research Center, Route du Jaurat, 1000, Lausanne, Switzerland}
 
\author{Elie Wandersman}
 \affiliation{Sorbonne Universit\'e, Centre National de la Recherche Scientifique, Laboratoire Jean~Perrin, LJP, F-75005 Paris, France.}

\author{Alexis M. Prevost}
\email{alexis.prevost@sorbonne-universite.fr}
 \affiliation{Sorbonne Universit\'e, Centre National de la Recherche Scientifique, Laboratoire Jean~Perrin, LJP, F-75005 Paris, France.}

\date{\today}
\begin{abstract}
 An experimental biomimetic tongue--palate system has been developed to probe human in-mouth texture perception. Model tongues are made from soft elastomers patterned with fibrillar structures analogue to human filiform papillae. The palate is represented by a rigid flat plate parallel to the plane of the tongue. To probe the behavior under physiological flow conditions, deflections of model papillae are measured using a novel fluorescent imaging technique enabling sub-micrometer resolution of the displacements. Using optically transparent newtonian liquids under steady shear flow, we show that deformations of the papillae allow determining their viscosity from 1~Pa.s down to the viscosity of water of 1~mPa.s, in full quantitative agreement with a recently proposed model [Lauga \textit{et al.}, \textit{Frontiers in Physics}, 2016, \textbf{4}, 35]. The technique is further validated for a shear--thinning and optically opaque dairy system.
\end{abstract}

\maketitle

%%%MAIN TEXT%%%%

\section*{Introduction}

While a key element of the eating experience, a detailed understanding of how humans perceive food texture currently remains elusive. The ability to discriminate accurately between small differences in food structure -- likely witnessed by anyone who has compared, for example full fat milk with semi-skimmed milk -- is well observed \cite{Engelen2005,TrulssonEssick}. From a food processing perspective, classic texture characterization techniques actually fail to reproduce the same resolution as reported by human subjects \cite{FatPerception,Nestec}. Consequently, efforts to replace fat and sugar, nutrients that also play a structural role in processed food, have only met with partial success \cite{stokes2013oral}, frustrating attempts to create food that is both more nutritionally balanced and also preferred by consumers. Furthermore, from a medical perspective, applying a better understanding of texture perception to dysphagia, a potentially life-threatening condition for sufferers, could lead to even more effective treatment options \cite{burbidge2016day}.
\newline
\indent Past work on characterizing physical texture in-mouth has frequently focused on approximating the tongue--palate system by smooth--walled parallel planes, between which a food matrix undergoes steady shearing flow\cite{kokini1977liquid}. This gave insight into the relevant shear rates and relative importance of competing physical effects such as bulk rheology and surface lubrication \cite{LEREVEREND201084}.
\newline
\indent The tongue is however not smooth but covered at its upper surface with both fungiform and filiform papillae. For humans, fungiform papillae are mushroom-shaped structures of typical diameter $\sim 400~\mu$m and height $\sim 100~\mu$m, and are distributed with a surface density\cite{ranc2006effect} of about 2~papillae/mm$^2$. They have been shown to be involved in texture perception. Using psychophysical experiments, the acuity of the oral mechanosensory system has indeed been linked to variations among tasters of their fungiform papillae density \cite{ESSICK2003289}. Experimental investigations in surface-averaged measurement devices have also evidenced that the fungiform-like topography did impact the tribological properties of the tongue--palate system\cite{ranc2006effect}. The role of filiform papillae for texture perception has however not been thoroughly studied. Filiform papillae are distributed over the tongue with a surface density\cite{ghom2008textbook} of about 5~papillae/mm$^2$. They are fibril-like structures with a typical length of about 250~$\mu$m and a radius of about 34~$\mu$m\cite{BookShimizu2012}, with no chemical receptors at their base~\cite{doty2015handbook}. Such high aspect ratios make them ideal candidates for mechanosensing \textit{via} fluid-structure interactions with liquids entering the oral cavity. Recently, theoretical investigations of isolated filiform papillae-like posts in the presence of viscous flows suggested that local deflections of a single papilla could generate stresses relevant for sensory input \cite{Lauga2016}. Such a mechanism for sensing stresses, while not yet causally shown, is hinted at in new observations in mice\cite{PaperPapillaePiezo2} which indicate that filiform papillae are co-located with nerve endings expressing Piezo2, a mechanosensitive ion channel whose involvement in touch mammalian cells has been evidenced recently\cite{Ranade2014}.
\newline
\indent Deflections of ciliated organs have been identified in mechanosensation processes for other biological systems\cite{Venier1994,weinbaum2010mechanotransduction,hudspeth2014integrating} and shown to provide an enhanced sensitivity that recent biomimetic systems have been able to reproduce, such as in the case of the fish lateral line\cite{Asadnia2016}. From this biomimetic principle, sensors have thus been developed, to measure liquid or air flow velocities near a boundary\cite{Bruecker2005,WexlerStone2013,PaekKim2014}. Within the context of in-mouth texture perception, no analog biomimetic approaches have however been reported so far.
\newline
\indent The present work is therefore dedicated to develop a novel biomimetic tongue--palate system and to investigate the local deflections of filiform papillae present in a shearing flow. We first detail the fabrication of the experimental setup. We then measure papillae's deflections over a large range of shear stresses and we provide a full experimental validation of the model of Lauga \textit{et al.}\cite{Lauga2016}. Finally, we discuss our results and the implications for the perception of texture of food products.

\section*{Materials and methods}
The functioning of the human tongue-palate system was mimicked by placing an elastomer based artificial tongue at the bottom of a rheometer cell on the one hand, and using the upper rigid rotating disk of the rheometer as an equivalent of the palate on the other hand.

\subsection*{Artificial tongues fabrication}

\begin{figure*}[!t]
	\centering
	\includegraphics[width=\textwidth]{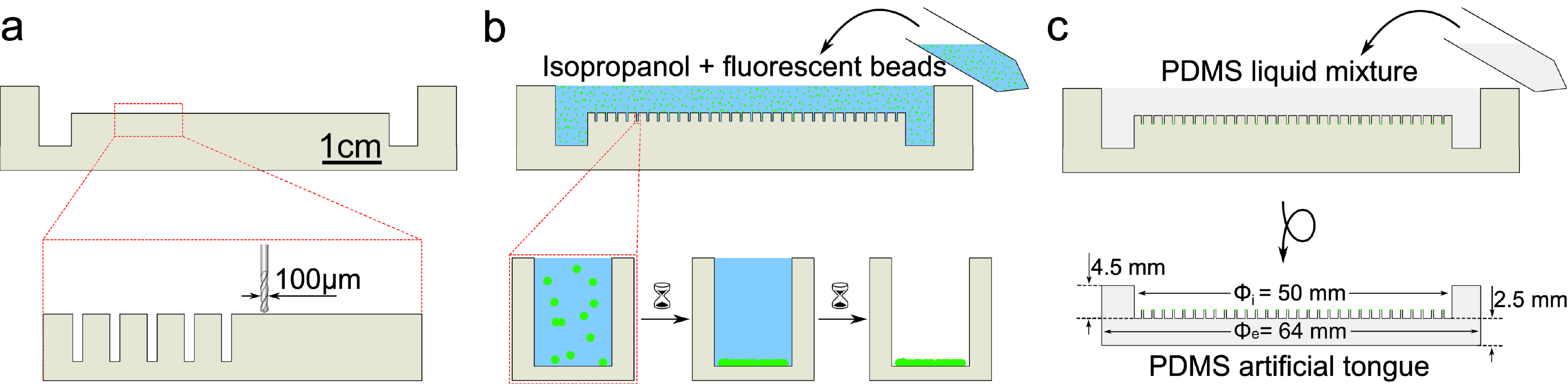}
	\caption{Sketch of an artificial tongue fabrication. (a) Sectional view along its diameter of the circular Plexiglas pool (upper panel). The lower panel shows the first step of the fabrication process where cylindrical holes with a diameter 100~$\mu$m and height 450~$\mu$m are microdrilled in the raised flat area. (b) Fluorescent beads of micrometric size dispersed in an isopropanol solution are deposited by sedimentation at the bottom of the cylindrical holes (upper panel, and lower panel for a close-up view). After sedimentation, the supernatant solution is pipetted and the pool is dried in an oven. (c) Eventually, a liquid PDMS/crosslinker mixture is poured in the pool and cured in an oven (upper panel). After careful unmolding, one is left with a transparent circular pool whose bottom is decorated with filiform-like papillae whose tips are covered with fluorescent beads (lower panel).}
	\label{fig:fig1}
\end{figure*}

\begin{figure}[h]
	\centering
	\includegraphics[width=0.48\textwidth]{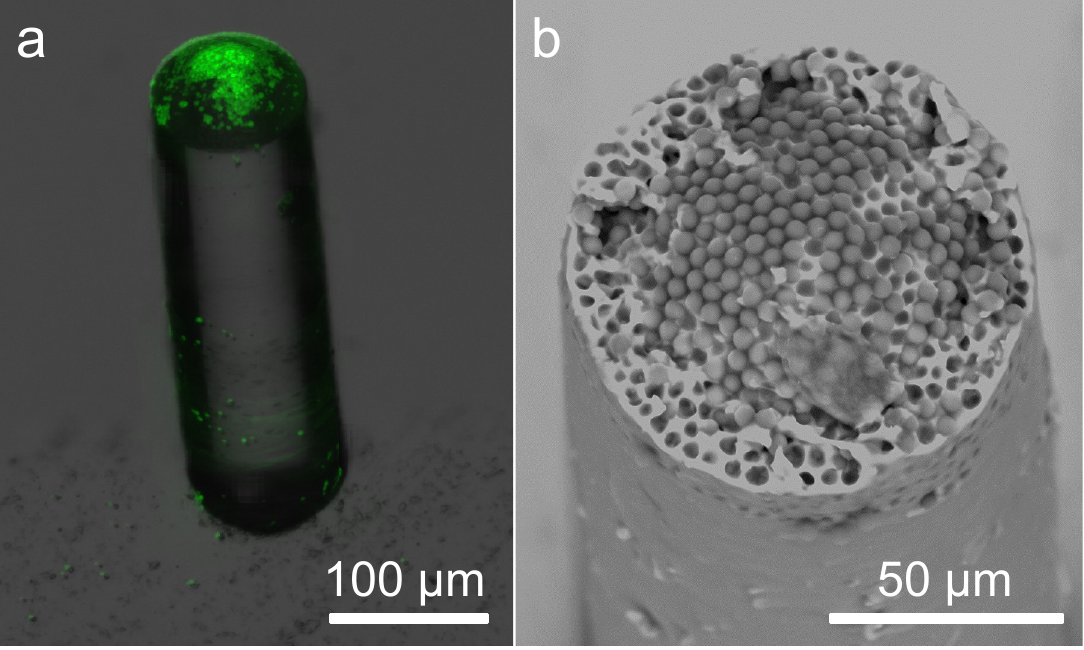}
	\caption{(a) Image of a single papillae whose tip is covered with fluorescent beads. This side view was obtained using a macroscope with fluorescence imaging capabilities. (b) SEM close-up view of the tip of a single papillae showing the existence of multiple layers of densely packed fluorescent beads.}
	\label{fig:fig2}
\end{figure}

Elastomer tongues were obtained with a combination of micromilling and molding techniques as sketched in Fig.~\ref{fig:fig1}. A Plexiglas mold shaped as a circular pool with a raised and polished bottom (\textit{see} Fig.~\ref{fig:fig1}a, upper panel, for a sectional view along its diameter), was first fabricated with an inner diameter $\Phi_i=50$~mm and an outer diameter $\Phi_e=64$~mm. Cylindrical holes of diameter 100~$\mu$m and depth 450~$\mu$m were then drilled in the raised bottom of the pool (Fig.~\ref{fig:fig1}a, lower panel) using a three axis commercial desktop CNC Mini-Mill machine (Minitech Machinary Corp., USA). These holes were distributed on a square grid with a 1~$\mu$m resolution, with a surface density of 10 holes/cm$^2$. With this density, hydrodynamic interactions between papillae are negligible\cite{PaperPapillaeDensity}. In a third step, an isopropanol solution containing green fluorescent beads of diameter [1--5]~$\mu$m (GFM, Cospheric, 1.3g/cc) at a concentration of 50~mg/L was poured onto the mold and left for 15~min at room temperature (Fig.~\ref{fig:fig1}b, upper panel). This duration was sufficient to allow sedimentation of beads at the bottom of the cylindrical holes. The supernatant solvent was then carefully pipetted and the pool dried for 15 min in a oven at $T=65$\textcelsius~(Fig.~\ref{fig:fig1}b, lower panel). Following this procedure, some beads could remain however adsorbed on the raised flat bottom of the pool. These unwanted beads were simply and efficiently removed using a commercial adhesive tape applied to the flat raised surface of the mold and carefully peeled off.
\newline
\indent  The artificial tongues were made of two optically transparent and commercially available cross-linked PolyDiMethylSiloxane (PDMS) elastomers, Sylgard 184 and Sylgard 527 (Dow Corning, USA), mixed in a 31:69 mass ratio. Prior to final mixing, both melts were first blended separately with their respective cross-linker agent, in a 10:1 stoichiometric ratio for Sylgard 184 and in a 50:50 ratio for Sylgard 527. Air bubbles present in the final mixture were removed with 10 min of centrifugation at 3000 revolutions/min followed by 2 min in a low pressure vacuum. This mixture was then poured into the Plexiglas mold (Fig.~\ref{fig:fig1}c, upper panel). Given the large aspect ratio of the micro-holes, air micro-bubbles could occasionally be trapped within the microstructures during pouring. These were removed by placing the mold filled with PDMS for an additional 10 min in a low pressure vacuum. The mixture was then cured for at least 12 hours in an oven at $T=65$\textcelsius. Finally, the polymerized tongue (with dimensions $\Phi_i=50$~mm, $\Phi_e=64$~mm, depth 4.5 mm, backing layer width 2.5 mm) was carefully unmolded to avoid rupture of the papillae (Fig.~\ref{fig:fig1}c, lower panel). We checked that papillae were intact and that the fluorescent beads were still present at their tip either using fluorescence microscopy (Fig.~\ref{fig:fig2}a) or Scanning Electron Microscopy (SEM) imaging (Fig.~\ref{fig:fig2}b). Note that in some cases, such as the one shown on Fig.~\ref{fig:fig2}a, a small number of beads could remain on the edges of the papillae. 

\subsection*{Tongues geometrical and mechanical characterizations}
As the beads deposition process can yield to variations in the final length $L$ of the papillae, we measured $L$ of all papillae considered in this work, either using optical microscopy or confocal imaging. For the former, $L$ was taken as the difference in height between the apex of the papillae and their base. Results of all combined measurements gave $L=435\pm7~\mu\mathrm{m}$.

\indent The Young's modulus $E$ of the artificial tongues was measured with JKR tests\cite{JKR1971,johnson1987contact} performed between an elastomer block ($50~\times~50$~mm, thickness 15~mm) prepared with the same ratio of Sylgard 184 and 527, and a planoconvex glass lens (optical grade, Melles-Griot 01LPX017, BK7, radius of curvature $R = 9.33$~mm). JKR tests were done with a custom made setup described elsewhere \cite{Prevost2013}. They consist in measuring the area of the circular contact between the lens and the elastomer as a function of the applied normal load, yielding a relationship from which $E$ can be deduced \cite{VerneuilPhDThesis2005}. Our measurements yielded $E = 0.80\pm0.16$~MPa. 

\subsection*{An experimental setup biomimetic of the oral cavity}

\begin{figure}[h]
	\centering
	\includegraphics[width=0.48\textwidth]{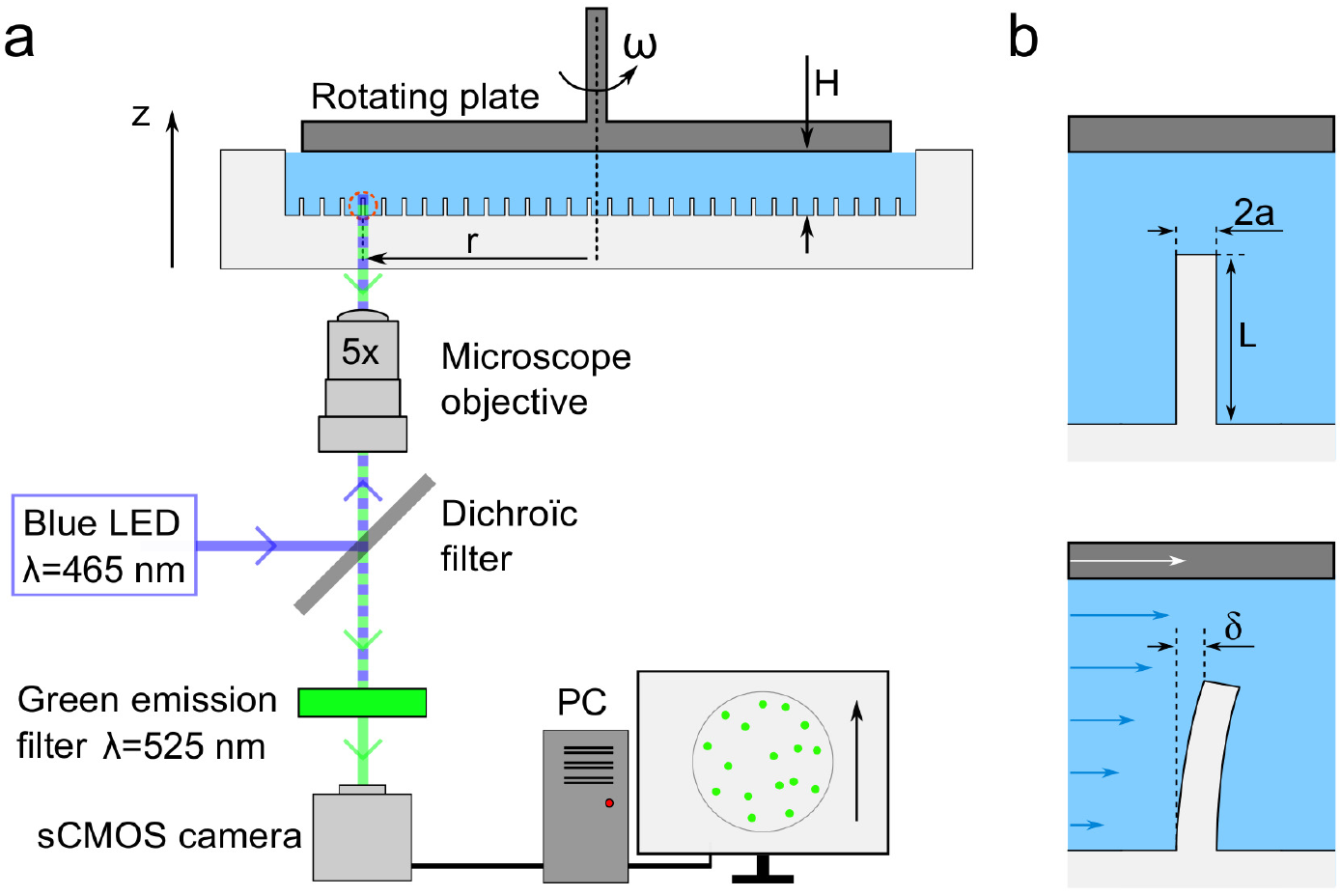}
	\caption{(a) Sketch of the tongue-palate biomimetic experimental setup (sectional view). The elastomer tongue is placed at the bottom of the rheoscope and filled with the liquid under study. The upper flat rotating plate of the rheoscope, playing the role of the palate, is brought in contact with the liquid at a distance $H$ from the base of the tongue. Papillae and their deflection are imaged in fluorescence with a 5$\times$~microscope objective and a fast camera. (b) Sketched sectional view of a single papilla at rest (upper panel) and submitted to a steady shear liquid flow (lower panel). Papillae are cylindrical with a diameter $2a$ and a total length from the base $L$. The deflection of their tip in steady state is noted $\delta$.}
	\label{fig:fig3}
\end{figure}

Artificial tongues were placed at the bottom of the cell of an Anton Paar MCR 302 rheo-microscope that combines both rheology measurements and imaging capabilities. Figure~\ref{fig:fig3}a shows a sketch of the experimental setup used for this work. Rather than moving the tongue as it is the case for humans, the artificial tongue is maintained at a fixed position in the frame of the laboratory, and the palate is moving. The palate's equivalent consists of a rigid and flat circular rotating plate (Anton Paar PP40, diameter 40 cm). For all experiments, the temperature was kept at $T=25$\textcelsius.

\indent Imaging of the deflections of papillae are performed in reflexion fluorescence microscopy using a long working distance microscope objective positioned directly beneath the artificial tongue (Fig.~\ref{fig:fig3}a, lower part). Only a small portion of the entire artificial tongue, that extends radially from the center of the cell to its perimeter, can be imaged. In practice, papillae under study are thus brought into this field of view by rotating the artificial tongue. The microscope objective is mounted on a manual translation stage that provides positioning in two directions, vertically and radially from the center of the rheometer's cell within the aperture. For our experiments, a 5$\times$~air objective (Edmund Optics, Plan-Apo) was used. Illumination is obtained with a high power blue LED (3W, $\lambda=465\pm5$~nm, Sodial(r)) whose beam is focused on the tip of the papillae. The light emitted by the fluorescent beads is collected with the combination of a dichroic filter and an emission filter (MD498 and MF525-39 respectively, Thorlabs, Inc.) onto a high resolution, fast and sensitive camera (sCMOS pco.edge 5.5, full resolution 2560$\times$2160~pixels, 16 bits) that can operate at a frame rate of 100~frames/s (fps) at full resolution. With the 5$\times$ objective, images have a resolution of 1.05~$\mu$m per pixel. In practice, only small portions of the images of size 200$\times$300~pixels$^2$, with a single papilla present in the field of view, were recorded at 100~fps.
\newline
\indent Finally, both the rheometer data and images were acquired simultaneously using a TTL trigger signal sent by the rheometer (controlled by its dedicated software Rheocompass 1.19, Anton Paar) to the camera, once rotation of the upper plate initiates.

\subsection*{Samples preparation}

In addition to Millipore deionized water ($\eta_0=1$~mPa.s), different water/glycerol solutions were used to calibrate the tongue-palate measurement system. These were prepared by mixing pure glycerol (Sigma-Aldrich) to Millipore deionized water at different mass ratio to obtain a logarithmic variation of the solutions dynamic viscosity $\eta$ ranging from 1~mPa.s to 1~Pa.s. For each glycerol solution, $\eta$ was measured with the rheo-microscope, operating at $T=25$\textcelsius~in a plate-plate geometry with the PP40 Anton Paar rotation plate, a 1~mm gap, and without the artificial tongue. It is known that glycerol solutions can be hygroscopic, but no significant change of viscosity was actually measured for the duration of the experiments.
\newline
\indent A model dairy product was also used in this work, consisting of a commercial semi-skimmed milk mixed with 0.5\% \textit{w}/\textit{w} xanthan gum hydrated overnight at room temperature. A small amount of sodium azide (Sigma-Aldrich) was added to prevent any bacterial development. The resulting liquid is non-Newtonian and possesses a rheological shear thinning behavior that was fully characterized with standard rheological measurements. In addition, it is completely opaque to visible light.

\subsection*{Modeling the deflection of an isolated papilla}
Following Lauga \textit{et al.} \cite{Lauga2016}, the tongue-palate system is modeled with two parallel plates separated by a gap $H$. The upper plate playing the role of the palate is rigid and moves laterally (with respect to the lower plate) at a velocity $U$. The lower plate mimicking the tongue is also considered rigid but covered with soft elastic cylinders of radius $a$ and length $L$, as equivalents of filiform papillae (Fig.~\ref{fig:fig3}b). In the absence of any fluid-structure hydrodynamic interactions between papillae, an assumption that is true in the limit of low surface density of papillae, Lauga and coauthors have derived a scaling relationship between the maximum deflection of the tip of a single cylindrical papilla subjected to a shear laminar liquid flow $\delta$  and the shear rate $\dot{\gamma}$. In steady state, $\delta$ reads
\begin{equation}
\delta = K\frac{L^5}{a^4} \frac{\eta}{E} \frac{U}{H} = \kappa \frac{\eta U}{H} ~\text{with}~\kappa= K\frac{L^5}{a^4 E}
\label{eq:laugamodel2}
\end{equation}
where $ K $ is a numerical factor whose value can only be determined experimentally. With the geometry of our experimental setup, one has $U=r \omega$, with $\omega$ the tool angular rotation velocity, $r$ the radial position of the papillae with the origin taken at the center of rotation of the upper tool.

\subsection*{Localizing the positions of papillae}
\begin{figure}[h]
	\centering
	\includegraphics[width=0.48\textwidth]{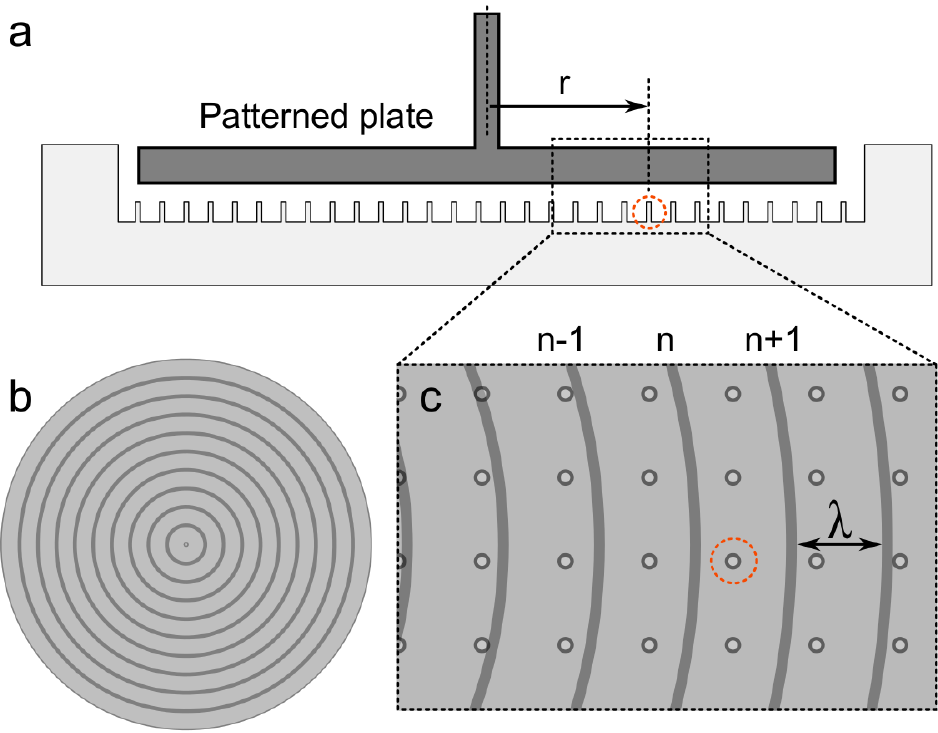}
	\caption{(a) Sketch of the experimental setup with the patterned rotating plate in place (sectional view). (b) Sketch of the lower surface of the patterned plate showing the concentric grooves (top view). (c) Principle of a papilla localization (close-up view): with the patterned plate in place, papillae imaged from below appear as sketched here. The papilla of interest (circled with the dashed red line) is localized with respect to its closest concentric circle (labelled here with the index $n$) whose radius is known by design.}
	\label{fig:fig4}
\end{figure}

Accurate papillae radial positions $r$ from the rotation axis of the rheoscope were determined as sketched on Fig.~\ref{fig:fig4}. A specific tool was designed in the form of a flat disk whose lower surface is engraved with a pattern of periodic concentric circular grooves of wavelength $\lambda=500~\mu$m (Figs.~\ref{fig:fig4}b-\ref{fig:fig4}c).

Once the artificial tongue is placed on the rheometer bottom cell, the patterned tool is inserted into the rheometer in place of the rotating tool, and brought (in air, \textit{i.e.} without any liquid) nearly in contact with the tip of the papillae. When imaged from below through the elastomer, one thus obtains an image of both papillae tips and concentric grooves as sketched in Fig.~\ref{fig:fig4}c. The position of the papilla of interest is then simply determined as the sum of the radius of the circular groove closest to the papilla and the distance of the papilla to the groove. The latter is obtained using image analysis, yielding a resolution on $r$ of typically less than 10~$\mu$m.

\subsection*{Setting the size of the tongue-palate gap $H$}
Accurate positioning of the upper plate was done in two successive steps. The first one consisted in making a full contact (without any liquid solution), between the upper tool and the surface of the bottom cell without the artificial tongue, to set a zero gap reference height. The second step consisted in positioning the artificial tongue on the bottom cell, still without any liquid solution, and lowering down progressively the upper tool, with 10~$\mu$m increments, until contacts occurred with the papillae tips. Contacts were identified by visualizing optically the displacement of papillae within the field of view, subsequent to a manual rotation of the upper plate. Knowing the zero gap reference height, the length of the papilla, and contacts heights allowed us to set $H$ to its desired value with a resolution of the order of $10\,\mu$m.

\subsection*{Papillae displacement measurements}

\begin{figure}[h!]
	\centering
	\includegraphics[width=0.48\textwidth]{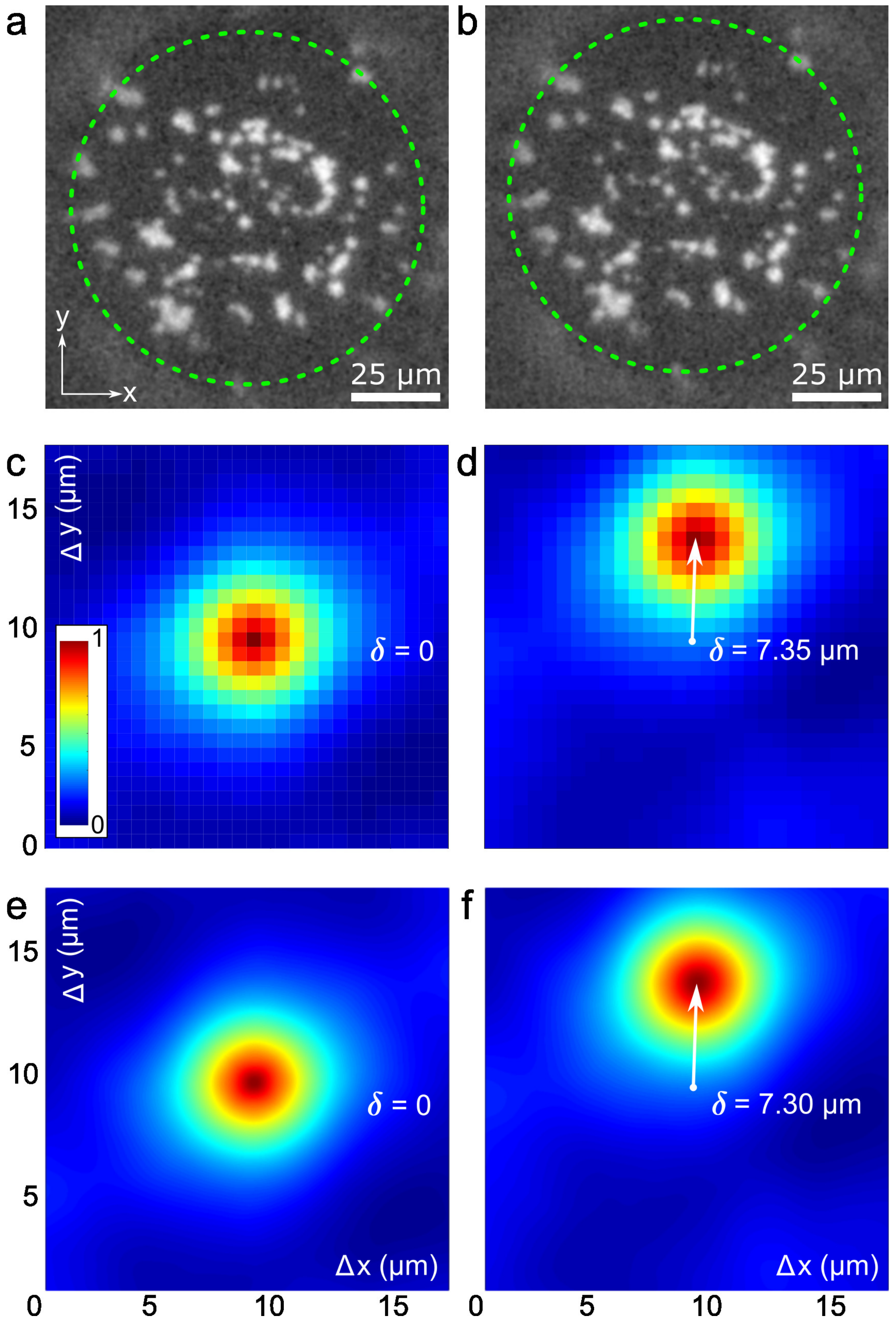}
	\caption{(a-b) Top view images in fluorescence of a single papilla in pure glycerol ($\eta=966$~mPa.s), (a) at rest and (b) when submitted to a shear circular flow ($r=12.64$~mm, $\omega=7.8$~rpm). The green dashed line circles delimit the perimeter of the papilla. The bright spots are the fluorescent microbeads embedded at the top of the papilla. (c-d) Pixel resolved normalized correlation function $C(\Delta x,\Delta y)$, (c) at rest and (d) in shear flow as a function of the shifts $\Delta x$ and $\Delta y$ along both $x$ and $y$ axes. Determination of its maximum yields the displacement vector $(\delta_x,\delta_y)$ of magnitude $\delta=7.35 \pm 1.05~\mu$m for the present case. (e-f) Subpixel resolved normalized correlation function $C(\Delta x,\Delta y)$ interpolated on a 100$\times$ finer mesh grid of the same papilla at rest (e) and in shear flow (f). Maximum of $C$ yields a displacement vector of magnitude $\delta=~7.30 \pm 0.10~\mu$m.}
	\label{fig:fig5}
\end{figure}

\begin{figure}[h]
	\centering
	\includegraphics[width=0.48\textwidth]{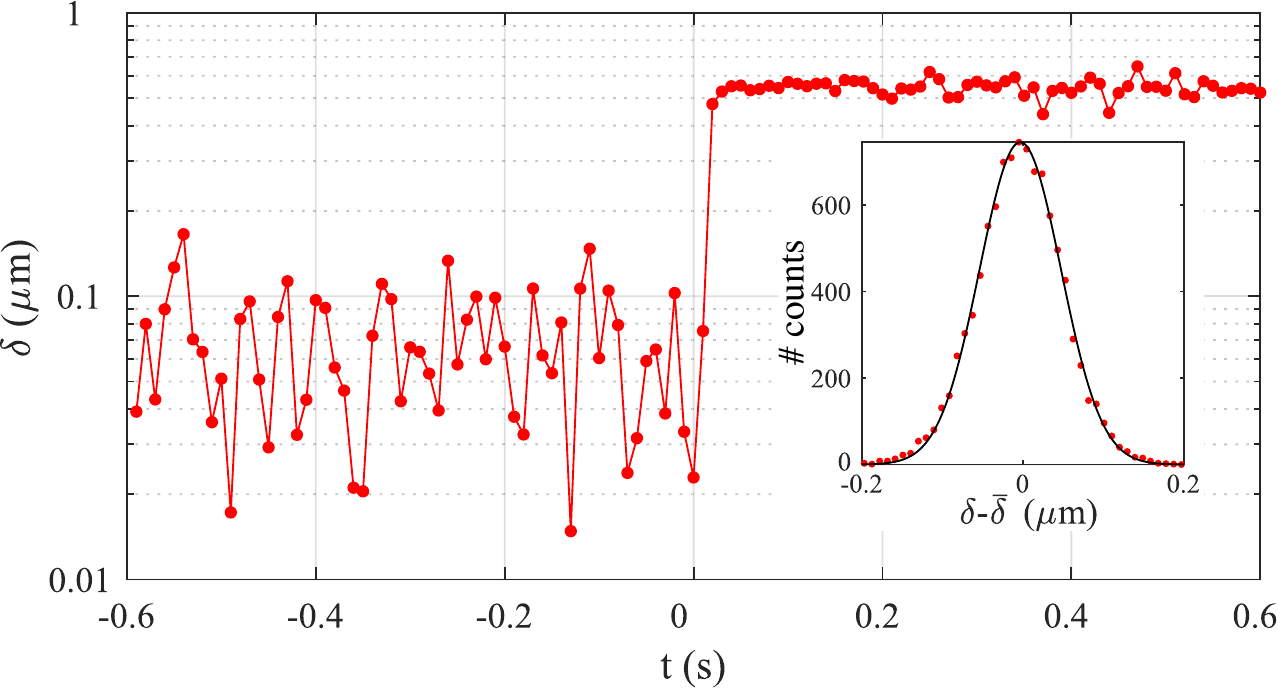}
	\caption{Semi-log plot of the displacement $\delta$ of a single papilla ($r=8.33$~mm) versus time during a typical experiment (water/glycerol 93.3\% w/w solution with $\eta=200~\pm~40$~mPa.s, $H=4$~mm, $\omega=0.44$~rpm, $\dot{\gamma}=0.10~\textrm{s}^{-1}$), after a sudden start of the rheometer tool at $t=0$~s. Inset: histogram of $\delta-\bar{\delta}$ where $\bar{\delta}$ is the mean of $\delta$ in steady state. The solid line is a gaussian fit of the histogram yielding a standard deviation of 71~nm.}
	\label{fig:fig6}
\end{figure}

Displacements of the extremity of a papilla were determined using Digital Image Correlation techniques. It consists in correlating images of the tip fluorescent bead markers when a flow is present, with a reference image of the papilla at rest (Fig.~\ref{fig:fig5}). Prior to this, all images were filtered with a Gaussian filter of standard deviation 1~pixel to remove high frequency noise. An Otsu's criteria automatic thresholding method was then used to set to a null value all pixels with intensity lower than the threshold. Both processes were done with Matlab R2017a (Mathworks, Inc., USA). The cross-correlation was performed in Fourier space using Matlab's built-in 2D FFT based algorithm, to yield the correlation function $C(\Delta x,\Delta y)$, where $\Delta x$ and $\Delta y$ are the shifts along both $x$ and $y$ axes of the image. Direct determination of its maximum provides the displacement vector $(\delta_x,\delta_y)$ (of magnitude $\delta=\sqrt{\delta_x^2+\delta_y^2}$) with one pixel resolution, \textit{i.e.} 1.05~$\mu$m with the 5$\times$~objective used here. Sub-pixel accuracy on the maximum location is commonly obtained by fitting $C(\Delta x,\Delta y)$ around its maximum with a 2D functional form, such as a polynomial or a gaussian surface for instance. For this work, we rather chose to interpolate $C(\Delta x,\Delta y)$ on a 100 times finer mesh, as described by Roesgen~\cite{Roesgen2003}. This interpolation method has the advantage of drastically reducing any pixel locking effect that is known to increase the error on the measured displacements. The method can however be computationally demanding, but the calculation time can be significantly reduced as shown by Guizar-Sicairos \textit{et al.} \cite{Guizar-Sicairos2008}, by limiting the interpolation to a region close to the maximum of $C$. Figure~\ref{fig:fig6} shows the results of such correlation based displacement measurements with a typical example of a single papilla that progressively bends from its initial rest position to its final position in steady flow. Taking into account all physical and numerical noises, the method was found to yield a resolution on the displacement $\delta$ of $\sim 71$~nm with an oversampling factor of 100. Such a resolution is in particular directly evidenced in the graph of Fig.~\ref{fig:fig6} in the steady state regime.

\indent Finally, note that the use of the correlation technique developed here is limited to small papillae deformations, \textit{i.e.} for $\delta/L < 10\%$. Indeed, for larger deformations, one observes shadow related masking and loss of focus of the fluorescent beads pattern that can significantly increase the error on the measured displacement.

\section*{Results and discussion}

\subsection*{Testing the response of a single papilla}

\begin{figure}[h]
	\centering
	\includegraphics[width=0.48\textwidth]{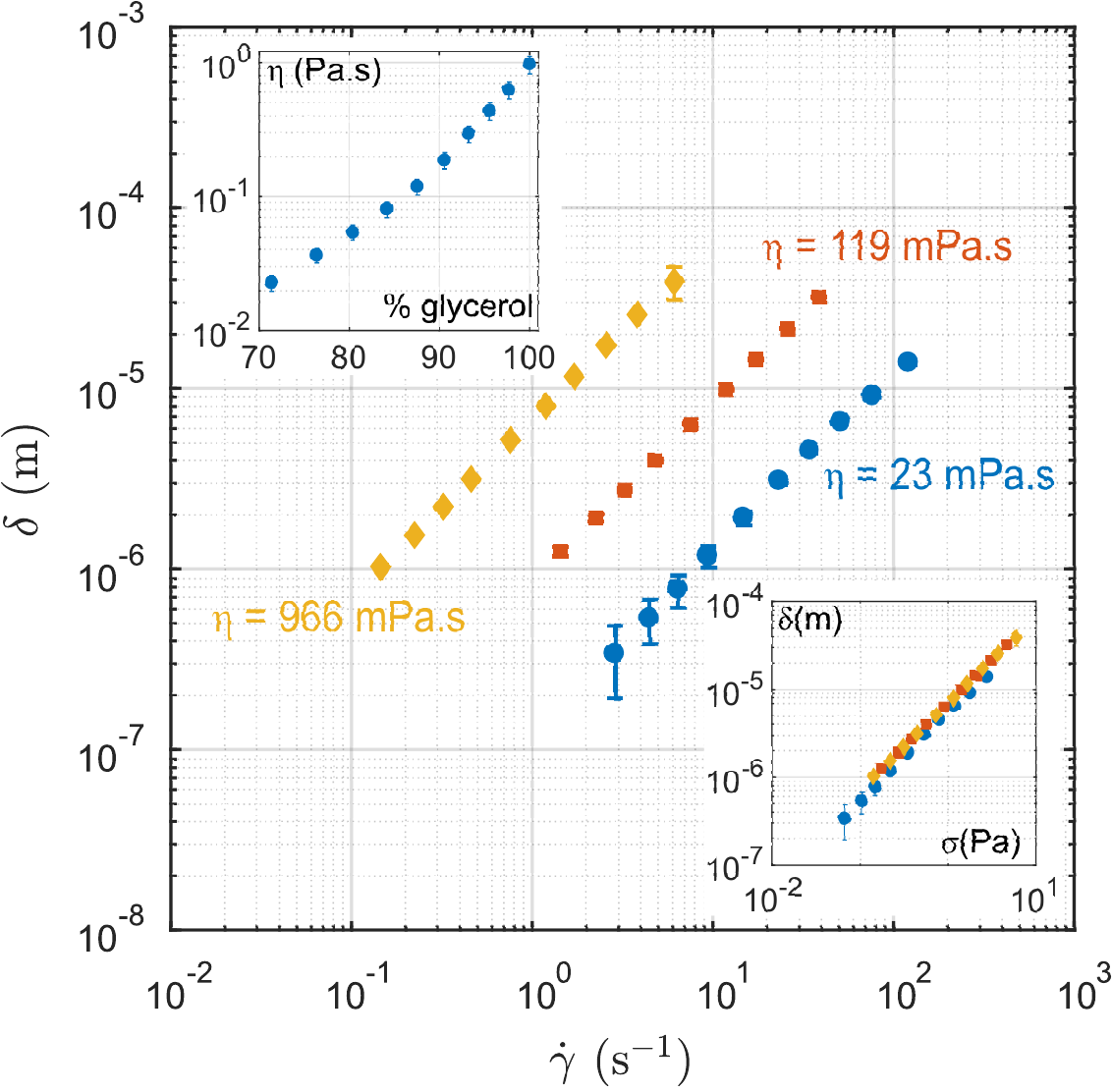}
	\caption{Displacement $\delta$ in steady state of the tip of one papilla as a function of the local shear rate $\dot{\gamma}=r \omega/H$ for three water/glycerol solutions with increasing dynamic viscosities, $\eta=23$~mPa.s (blue disks), $\eta=119$~mPa.s (red squares) and $\eta=966$~mPa.s (orange diamonds). For these experiments, $H=4$~mm and $\omega$ was varied. The same papilla was used but its position $r$ could vary slightly when changing the solution due an inherent tongue repositioning ($r=12.6, 13.1, 13.3$~mm for $\eta=23, 119, 966$~mPa.s respectively). Upper inset: $\eta$ as a function of the water/glycerol mass ratio. Lower inset: $\delta$ as a function of $\eta r \omega /H$ for all three $\eta$.}
	\label{fig:fig7}
\end{figure}

We first performed a series of experiments using newtonian liquids of different viscosities to test the validity of the linear scaling prediction of Eq.~\ref{eq:laugamodel2} (\textit{see} section "Materials and methods") that relates $\delta$ to $\eta$ and $\dot{\gamma}=r \omega/H$ in the steady state flow regime. This was done using different water/glycerol solutions with the same artificial tongue and the exact same papilla positioned at $r<\Phi_i/2-H$ to avoid any hydrodynamic boundary effect. Note that the radial position of the papilla $r$ could change slightly between experiments due to an inherent manual repositioning of the tongue, and it was thus systematically determined as explained earlier (\textit{see} section ``Materials and methods -- Localizing the positions of papillae''). The water/glycerol solutions were prepared according to the protocol described in the Materials and methods section, with $\eta$ ranging from 10$^{-2}$~Pa.s to 1~Pa.s (\textit{see} Fig.~\ref{fig:fig7}, upper inset, for the measured relationship $\eta$--water/glycerol mass ratio).

\indent The experiments consisted in first accurately positioning the upper plate at a distance $H=4$~mm above the base of the papillae, and second in pipetting a volume of 7.5 mL of the water/glycerol solution within the artificial tongue. The upper plate was then set into rotation at a constant angular velocity $\omega$, causing the papillae to  deflect. The steady state flow regime and thus papillae's maximum bending was typically attained in less than 1~s. In practice, each experiment consisted of 11 successive 10~s long measurements. The first one was performed without any flow (papilla at rest) to provide an unperturbed reference state, while the 10 subsequent measurements were done with increasing $\omega$. Analysis of the displacements were performed in the steady state regime.
\newline
\indent The main panel of Fig.~\ref{fig:fig7} shows for a given papilla the results of these experiments with $\delta$ as a function of $\dot{\gamma}$ for three distinct viscosities, $\eta=23$~mPa.s, $\eta=119$~mPa.s and $\eta=966$~mPa.s. Each single point corresponds to a single measurement and its error bar is taken as the standard deviation of $\delta$ over the 10~s duration of the experiment in steady state. For all three viscosities shown here, $\delta$ is found to be proportional to $\dot{\gamma}$ over almost two decades in $\dot{\gamma}$, in agreement with the prediction of Eq.~\ref{eq:laugamodel2}. In addition, all three curves can be rescaled on the same master curve once the axis is multiplied by $\eta$, as shown on the lower inset of Fig.~\ref{fig:fig7}. At a fixed shear rate, $\delta$ is thus directly proportional to the local viscosity $\eta$ of the probed solution over almost two decades in $\eta$.

\begin{figure}[h]
	\centering
	\includegraphics[width=0.48\textwidth]{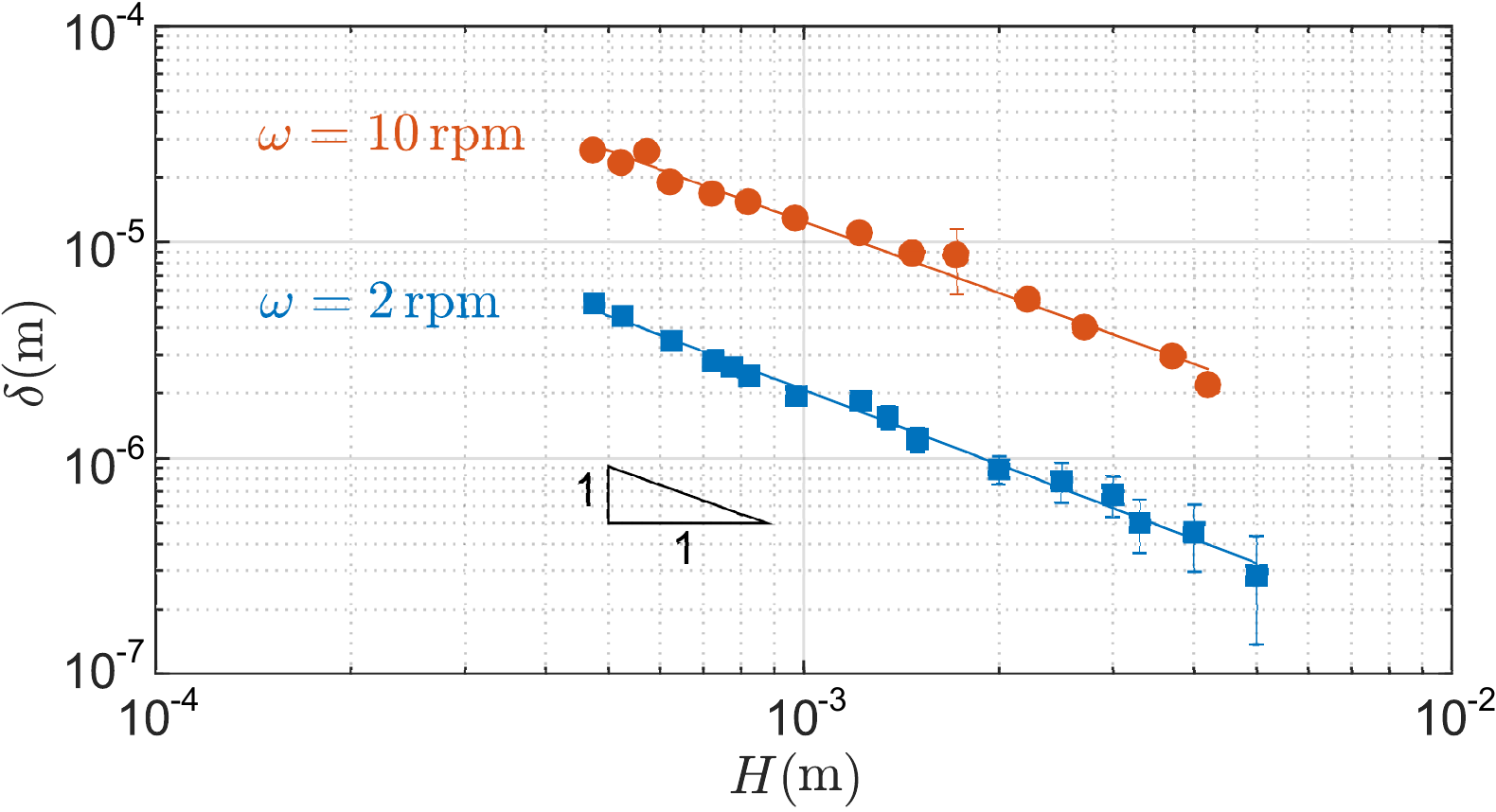}
	\caption{Displacement $\delta$ in steady state of a single papilla ($r=14.83$~mm) as a function of $H$ for two different angular velocities $\omega$, in pure glycerol ($\eta=459\pm62$~mPa.s as measured by the rheometer). The dashed lines are power law fits to the data of the form $A H^{\alpha}$. For $\omega=2$~rpm, the fit yields $A=0.32\pm0.02$ and an exponent $\alpha=-1.15\pm0.06$. For $\omega=10$~rpm, the fit gives $A=1.10\pm0.03$ and $\alpha=-1.10\pm0.09$.}
	\label{fig:fig8}
\end{figure}

The experiments we have considered so far were limited to the case where $H=4$~mm is much larger than the typical size $L$ of the papillae. In the last stages of food texture perception however, \textit{i.e.} immediately prior to swallowing, the tongue can approach significantly the palate (and even eventually be in close contact), and $H$ can be of the order of $L$. We have thus also investigated how $\delta$ varied with $H$ for a single papilla. This was done with a water/glycerol solution of viscosity $\eta=459\pm62$~mPa.s, at two different angular velocities $\omega$. As shown in Fig.~\ref{fig:fig8}, for both $\omega$, $\delta \propto 1/H$ for $H \in$ [0.5--5]~mm. These measurements fully validate the predictions of Eq.~\ref{eq:laugamodel2}, even when $H \approx L$.

\subsection*{Comparing the response among papillae}

\begin{figure}[ht]
	\centering
	\includegraphics[width=0.48\textwidth]{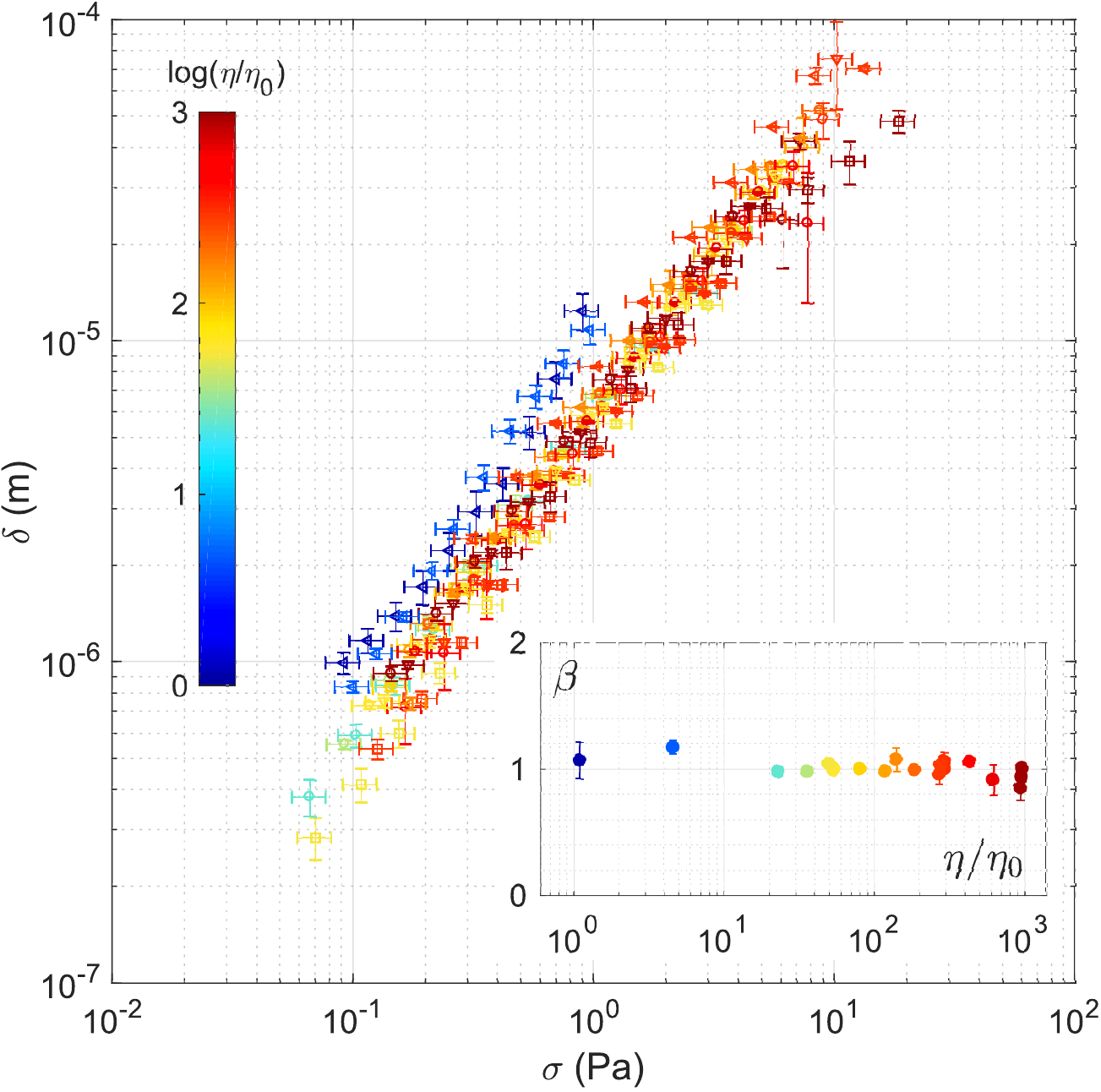}
	\caption{Log-log plot of the displacement $\delta$ in steady state versus $\sigma=\eta \dot{\gamma}$. Each color corresponds to one viscosity $\eta$. Symbols correspond to different experiments. All data were fitted with a power law of exponent $\beta$, not shown on this graph for clarity. Inset: Exponent $\beta$ as a function of $\eta/\eta_0$ where $\eta_0=10^{-3}$~Pa.s is the viscosity of water.}
	\label{fig:fig9}
\end{figure}

Our measurements have focused so far on the response of a single papilla to changes in viscosity. Additional experiments were thus performed to see how robust the measured linear response behavior pertains among papillae located at different radial positions and across different artificial tongues. For this, three different papillae on the same artificial tongue were considered, as well as one papilla on another tongue. Similarly as before, the papillae were chosen sufficiently far away from the boundaries of the tongue ($r<\Phi_i/2-H$). Each series of experiments were not only done with the exact same protocol as before, but were also repeated 3 times with the same papilla, without any repositioning of the tongue. In addition, for one viscosity, these measurements were repeated 3 times by changing the solution. A total of 9 measurements were thus obtained per angular velocity $\omega$. In addition to the 10 water/glycerol solutions used before, Millipore deionized water ($\eta_0=1$~mPa.s) and a 45/55 \textit{w}/\textit{w} mixture of water/glycerol ($\eta=4$~mPa.s) were used for these experiments. Figure~\ref{fig:fig9} shows the results of all experiments combined in a log-log scale, with $\delta$ as a function of the shear stress $\sigma=\eta \dot{\gamma}$ varying over nearly three orders of magnitude (and with $\eta$ varying exactly over three orders of magnitude from 10$^{-3}$~Pa.s to 1~Pa.s). For each point at a fixed $\omega$, error bars were taken as the standard deviation over all 9 measurements performed. Further checks of the linearity of Eq.~\ref{eq:laugamodel2} were obtained by fitting data points at a fixed $\eta$ with a power law of exponent $\beta$ (for sake of clarity, fits are not displayed on the figure). As shown in the inset of Fig.~\ref{fig:fig9}, for all viscosities, $\beta$ is found to be very close to unity, in full agreement with Eq.~\ref{eq:laugamodel2}. Note that the blue colored points on Fig.~\ref{fig:fig9} are slightly shifted upwards with respect to the other colored symbols. The blue-colored points do correspond to measurements obtained with a different tongue, in contrast with the other points obtained with different papillae on the same tongue. For this new tongue, while the linear behavior is still measured, the value of the $\kappa$ parameter (\textit{see} Eq.~\ref{eq:laugamodel2}) is slightly larger. We checked that the values of both $L$ and $a$ were unchanged within experimental errors. This measured shift is thus likely due to variations of the apparent elastic modulus of the papillae, presumably due to the fluorescent particles inclusion process. For absolute measurements purposes, such small scatter thus implies that a calibration has to be systematically performed for each new tongue.

\subsection*{Testing the response of a food related complex fluid}
\begin{figure}[ht]
	\centering
	\includegraphics[width=0.48\textwidth]{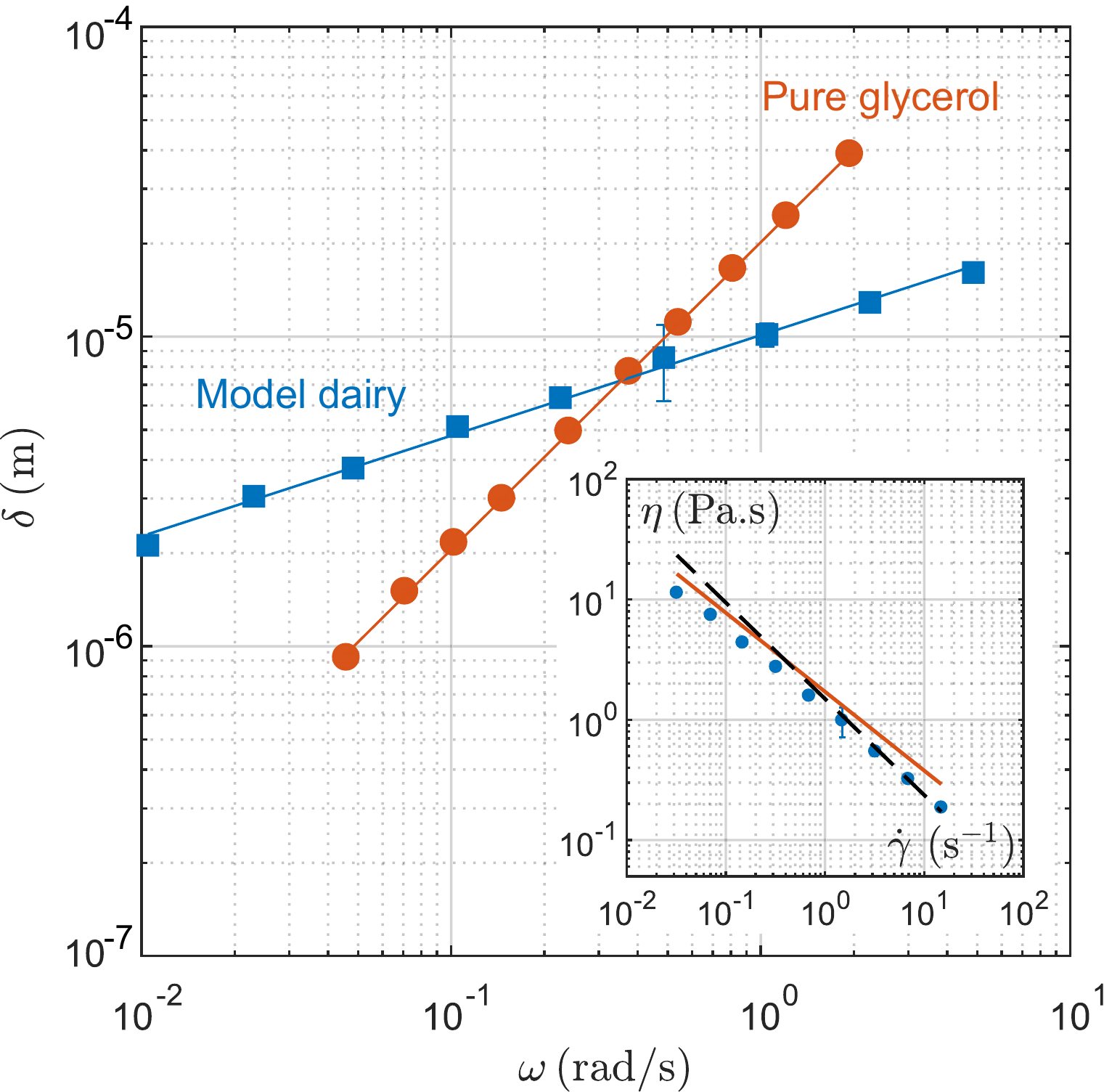}
	\caption{Log-log plot of the displacement $\delta$ in steady state versus $\omega$ for both a pure glycerol solution (blue squares) and the model dairy product (red disks). Each point corresponds to a single experiment. Thick solid lines are fits to the data with a power law $A \omega^\beta$. For the glycerol solution, the fit yields $A=20.2\pm0.4~\mu$m/rad.s$^{-1}$ and $\beta=0.99\pm0.01$. For the model dairy, $A=10.1\pm0.5~\mu$m/rad.$^{-1}$ and $\beta=0.32\pm0.02$. Inset: Log-log plot of $\eta$ versus $\dot{\gamma}$ for the model dairy (blue solid disks) compared to both the viscosity measured with the rheometer (red solid line) and the measurements of Zatz \textit{et al.}\cite{ZatzKnapp1984} (black dashed line).}
	\label{fig:fig10}
\end{figure}

So far, all of our previous measurements were performed with simple liquids that are optically transparent. A food related product, such as a dairy product for instance, is however made of submicrometric to micrometric particles that scatter visible light and can make it completely opaque. We thus also tested our imaging setup with a model dairy product (\textit{see} section ``Materials and methods -- Samples preparation'' for details) sheared between the artificial tongue and the palate (Fig.~\ref{fig:fig10}). The rheological properties of this product have also been measured in plane-plane geometry (inset of Fig.~\ref{fig:fig10}, red solid line). Shear thinning behavior was evidenced, in agreement with previous measurements\cite{ZatzKnapp1984} (inset of Fig.~\ref{fig:fig10}, black dashed line).

Thanks to its reflexion based imaging through the cylindrical papillae, measurements of the bending of papillae in steady flow were easily achieved with the same resolution on displacement as for optically transparent liquids. Figure~\ref{fig:fig10} shows the results of such measurements with $\delta$ as a function of the angular velocity $\omega$, for the only papilla that was probed, and with a single experiment per angular velocity. For comparison, the same papilla was also used with a pure glycerol Newtonian solution. For this glycerol solution, one recovers the expected linear behavior contained in Eq.~\ref{eq:laugamodel2}, as shown in Fig.~\ref{fig:fig10} with the power law fit of the data of exponent $\beta \sim 1$. For the model dairy product however, one clearly measures a drastic deviation from the linear behavior, characteristic of a shear thinning rheological behavior. Indeed, $\delta$ is found to vary with $\omega$ as a power law of exponent $\beta \sim 0.3$, over almost three decades in $\omega$.  

\subsection*{Discussion}
Several key aspects of the current work can be drawn. First of all, classical methods to manufacture micro-pillars of moderate aspect ratios (\textit{i.e.} typically not more than 10) usually involve molding of elastomers either in resin based substrates fabricated with microphotolithography processes or in sacrificial wax based templates \cite{Bruecker2005}. In our work, we have used a micro-drilling based technique of a plastic Plexiglas sample that provides micrometric accuracies in all 3 directions, that is simple, easy to implement and highly reproducible. 
\newline
\indent To follow the displacement of the tip of the pillar, the classical approach is to embed the extremity of the pillar with a marker either in the form of a reflective coating\cite{Bruecker2005,PaekKim2014} or in the form of a fluorescent dye such as rhodamine, as done recently\cite{Bruecker2016}. An alternative method, that does not require any tip marking, has also been implemented and consists in using the optical properties of transparent PDMS based cylindrical pillars that act as wave guides\cite{Bruecker2005}. Illumination light propagates from the base of the pillars to their extremity where it is focused, yielding a bright spot that can be detected optically. The same wave guide effect has been reported by Bruecker\cite{Bruecker2016}, with fluorescent light of tips covered with rhodamine. In this case, the fluorescent light is both emitted at the tip and at the base of the pillar as the primary fluorescence light propagates down to the base with the wave guide effect. In our work, we propose an alternate method where fluorescent microbeads are embedded by sedimentation in the plastic mold and then cast within PDMS at the top of the cylindrical pillars. This method avoids any possible diffusion of the fluorescent markers\cite{Bruecker2016} and yields a robust and reusable marking mechanism. Compared to the wave guide mechanisms for which flexion of the pillars lowers the intensity of the reflected light from the tip\cite{PaekKim2014}, our method offers stable intensity patterns that are independent of the amplitude of the deformation. Moreover, it can be used regardless of the optical index mismatch between the surrounding fluid and the pillar. 
\newline
Once the tip has been marked, image analysis methods are generally used to locate the position of the tip. These can rely either on morphological analysis\cite{Bruecker2005} and image auto-- and cross-- correlation techniques\cite{Bruecker2005,Bruecker2016}, that can yield a subpixel resolution on the measured displacements. In our case, we have used fluorophores in the form of microbeads embedded at the extremity of the pillars. They form a well defined intensity pattern that increases the resolution of the correlation technique. Moreover, the minimum measurable displacement is only set by the noise of the correlation method, in contrast with the work of Bruecker\cite{Bruecker2016}, for which a minimal displacement of the order of the pillar radius is required. Last, the use of an interpolation based method\cite{Roesgen2003} to locate with subpixel accuracy the location of the tip position, rather than using functional fits of the 2D correlation function, avoids any pixel biased effects and reduces significantly computational time\cite{Guizar-Sicairos2008}.
\newline
\indent Second of all, these marking and image correlation methods allow us to probe the deflection of the tip with a spatial resolution of about 70~nm, for deflections up to several tens of micrometers. We were thus able to probe the theoretical predictions of Lauga and coauthors\cite{Lauga2016} (\textit{see} Eq.~\ref{eq:laugamodel2}) in a steady flow on two orders of magnitude in the tip displacement $\delta$. This was done by varying both the viscosity on three orders of magnitude and the shear rate on two orders of magnitude. To the best of our knowledge, exploring this relationship over such a large range of shear stress has never been reported and thus constitutes a rigorous validation of the theoretical model of Lauga \textit{et al.}\cite{Lauga2016}. Indeed, even though numerous linear relationships between the deflection in steady state and the fluid velocity have been reported\cite{Venier1994,Bruecker2005,WexlerStone2013}, no systematic variation of viscosities has been performed. Our work demonstrates that the artificial papillae system can be used as a sensitive microrheometer to measure viscosities in a way similar to that operated by the tongue--palate system.
\newline
\indent Finally, interestingly, the use of fluorescent microbeads as markers enables measurements of optically opaque liquids such as dairy food products. Our system has indeed been successfully used to probe rheological properties of yogurts and chocolate mousses (not shown). As a model dairy, we have used a mixture of milk and xanthan that exhibits, as many food products, a shear thinning behavior. We show that in this case the deflections of papillae are weakly dependent on the shear rate. From a biological point of view, our measurements thus imply that the sensory input, which is most likely related to the bending of papillae, is weakly dependent on the in-mouth shear stress. Texture perception of food products appears therefore robust across varying chewing conditions. Such a feature echoes tactile perception mechanisms in other contexts. In human tactile digital perception for instance, it is known that such robustness (to variations in the finger/object frictional properties for example...) provides the human hand the ability to maintain a stable grasp at all times\cite{FlanaganJohansson2009}. Similarly, for rodents who use their facial whiskers to detect objects in their immediate vicinity, the whiskers vibrations elicited upon contact provide a detection mechanism that is robust over a large range of exploration conditions \cite{Claverie2017}. 
The present work suggests that the in-mouth texture perception of food products that have shear thinning rheological properties is equally robust to variations in the exploratory chewing conditions.

\section*{Conclusion and perspectives}
In this work, we have developed and calibrated a novel sensor biomimetic of the human tongue--palate system, based on the deflections of cylindrical soft asperities reminiscent of the filiform papillae. We have shown that their deflection was proportional to the applied shear stress. We have provided a rigorous validation of the elastohydrodynamic model developed by Lauga and coauthors. For future work, two main aspects will be addressed. First, the present experiments have focused on shear induced papillae deflections. It is known however from psychophysical studies\cite{Whillis1946} and ultrasound imaging of oral manipulation of fluids \cite{DEWIJK20031} that the in-mouth texture perception also involves squeeze flows corresponding to an upward motion of the tongue, whose consequences on papillae deflections have been theoretically modeled by Lauga \textit{et al.}\cite{Lauga2016}. Experimental investigations of this exploration condition remain to be performed. Second, in this work, we have only considered the case of tongues covered with a low surface density of papillae, for which hydrodynamics interactions between papillae are negligible. Human tongues however, possess a much higher density of filiform papillae ($\sim 5$~papillae/mm$^2$), at which non-trivial coupled elastohydrodynamic effects are likely to occur \cite{AlvaradoHosoiNaturePhysics2017}. The effect of papillae surface density on their deflections under shear will be the core of a forthcoming paper that will address them both experimentally and theoretically\cite{PaperPapillaeDensity}. This work is of relevance to the psychophysics community since it will also help to better understand the textural coding of complex fluids in the oral cavity and allow to design future psychophysics studies associated with their soft matter physics interpretation. \\

\section*{Acknowledgments}
Fruitful discussions with E.~Lauga (DAMTP, University of Cambridge, Cambridge, UK) are greatly acknowledged. We also thank the imaging platform of IBPS (Sorbonne Universit\'e, Paris, France) for the macroscope measurements, and L.-L.~Pontani (LJP, Sorbonne Universit\'e, Paris, France) for her help in confocal microscopy imaging of the papillae. Finally, we are grateful to A.~Chateauminois (SIMM, ESPCI, Paris, France) for giving us access to the SEM facilities of IPGG (Paris, France).
%
%%%%END OF MAIN TEXT%%%

%\balance


%%%REFERENCES%%%
\bibliography{biblio} %You need to replace "rsc" on this line with the name of your .bib file

\end{document}